\documentclass[a4paper]{jpconf}

\usepackage{graphicx}
\usepackage{amsmath,amssymb,amsfonts,amscd}

\newcommand{\msol}{M_{\odot}}

\newcommand{\zehnh}[2]{{#1} \times 10^{#2}}

\newcommand{\msec}{\textrm{ms}}
\newcommand{\km}{\textrm{km}}
\newcommand{\cm}{\textrm{cm}}

\newcommand{\sek}{\textrm{s}}
\newcommand{\isek}{\textrm{s}^{-1}}

\newcommand{\gccm}{\textrm{g\,cm}^{-3}}

\newcommand{\Gauss}{\textrm{G}}

\newcommand{\first}{$1^{\mathrm{st}}$}
\newcommand{\second}{$2^{\mathrm{nd}}$}
\newcommand{\third}{$3^{\mathrm{rd}}$}
\newcommand{\nth}[1]{${#1}^{\mathrm{th}}$}

\newcommand{\mri}{\textrm{MRI}}
\newcommand{\MMM}{\mathcal{M}_{r\phi}}
\newcommand{\MMMte}{\mathcal{M}_{r\phi}^{\mathrm{term}}}

\newcommand{\figref}[1]{Fig.\,\ref{#1}}
\newcommand{\tabref}[1]{Tab.\,\ref{#1}}

\newcommand{\jcop}{J. Comput. Phys.}

\newcommand{\apj}{ApJ}
\newcommand{\mnras}{MNRAS}

\newcommand{\aap}{A{\&}A}

\newcommand{\apjl}{ApJL}

\begin{document}

\title{Termination of the MRI via parasitic instabilities in
  core-collapse supernovae: influence of numerical methods}

\author{T.~Rembiasz${}^{1}$, M.~Obergaulinger${}^{1}$, 
  P.~Cerd{\'a}-Dur{\'a}n${}^{1}$, M.{\'A}.~Aloy${}^{1}$ and E.~M{\"u}ller${}^2$}

\address{
  ${}^1$ Departamento de Astronom\'{\i}a y Astrof\'{\i}sica,  Universidad de Valencia,  C/ Dr.~Moliner 50, 46100 Burjassot, Spain
  \\
  ${}^2$ Max-Planck-Institut f{\"u}r Astrophysik, Karl-Schwarzschild-Str.~1, 85748 Garching, Germany
}

\ead{martin.obergaulinger@uv.es}

\begin{abstract}
  We study the influence of numerical methods and grid resolution on
  the termination of the magnetorotational instability (MRI) by means
  of parasitic instabilities in three-dimensional shearing-disc
  simulations reproducing typical conditions found in core-collapse
  supernovae.  Whether or not the MRI is able to amplify weak magnetic
  fields in this context strongly depends, among other factors, on the
  amplitude at which its growth terminates. The qualitative results of
  our study do not depend on the numerical scheme.  In all our models,
  MRI termination is caused by Kelvin-Helmholtz instabilities,
  consistent with theoretical predictions.  Quantitatively, however,
  there are differences, but numerical convergence can be achieved
  even at relatively low grid resolutions if high-order reconstruction
  methods are used.
\end{abstract}

\section{Introduction}
\label{Sek:Intro}

The iron core of a massive star (zero-age main-sequence mass
$M_{\mathrm{ZAMS}} \gtrsim 8 \, \msol$) collapses within a few hundred
milliseconds from a radius of a few thousand kilometer and a central
density of the order of $\rho_{\mathrm{c}} \sim 10^{10}\, \gccm$ to a
proto-neutron star (PNS) with a radius of $\sim50\,$km and a central
density above nuclear matter saturation density, $\rho_{\mathrm{nuc} }
\sim \zehnh{2}{14}\, \gccm$.  A hydrodynamic shock wave forms when the
core bounces because of the stiffness of nuclear matter. This prompt
shock stalls because of severe energy losses due to
photo-disintegration of heavy nuclei into free nucleons, and turns
into an accretion shock initially located at a radius $\sim 100\,$km,
through which matter continues to accrete onto the PNS.  Most of the
gravitational binding energy liberated during the collapse is radiated
away in the form of neutrinos. A small fraction of these neutrinos
inteact with matter in the post-shock layers depositing energy, which
revive the stalled shock and eventually cause a core-collapse
supernova explosion (CCSN). A failure to revive the shock before
accretion increases the mass of the PNS beyond the maximum mass that
can be supported against gravity, will lead to a secondary collapse to
a black hole.  Hence, the fate of the core hinges on the efficiency of
the energy transfer to the post-shock matter, which may be enhanced by
various hydrodynamic instabilities (convection and the standing
accretion shock instability, SASI).  The exact conditions for the
neutrino-driven explosion mechanism to launch an explosion of the star
compatible with the observations (in particular concering the
explosion energy) is currently a matter of intense investigation (for
a review see, e.g., \cite{Mezzacappa2005},
\cite{Janka__2012__ARNPS__ExplosionMechanismsofCore-CollapseSupernovae},
\cite{Kotake2012}, \cite{Burrows2013}).

If the pre-collapse core rotates rapidly, part of its rotational
energy might be tapped to aid the neutrino-driven mechanism.  The
efficiency of this effect is greatly enhanced in strongly magnetised
cores.  However, the presence of both rapid rotation and strong fields
is unlikely except for possibly a very small set of progenitors.
Therefore, assuming a rapidly rotating core, the prospects for
rotationally aided or even rotationally driven explosions greatly
increase if strong magnetic stresses can be generated after collapse,
i.e.~if a weak seed field is efficiently amplified during the time
before either an explosion is launched by other mechanisms or the PNS
collapses to a black hole.

Angular momentum conservation during collapse causes the core to speed
up and strong differential rotation is generated at the edge of the
homologously collapsing inner core, even for initially rigidly
rotating iron cores.  This opens additional channels for the
amplification of seed fields besides flux conservation in the
(compressive) accretion flow and a possible turbulent dynamo driven by
hydrodynamic instabilities that do not rely on rotation.  Winding of a
pre-existing poloidal field component creates a toroidal component
that grows linearly with time
(e.g. \cite{Meier_etal__1976__ApJ__MHD_SN}).  Possibly more important
still, a significant fraction of the core fulfils the instability
criterion for the magneto-rotational instability (MRI)
\cite{Akiyama2003}. In its most basic form, MRI requires a negative
radial gradient of the angular velocity, $\partial_{r} \Omega < 0$,
and a weak background field to increase field perturbations on an
exponential timescale. The growth time of the instability is
comparable to the rotational period of the core that can be much
shorter than the post-collapse evolution timescale ($\sim 100\,$ms).

The MRI has been intensively studied in the context of accretion
discs, where it is considered to provide the most important source of
turbulence and enhanced angular-momentum transport
\cite{Balbus_Hawley__1998__RMP__MRI}.  In CCSN, indications for its
activity have been found in global axisymmetric simulations
\cite{Obergaulinger_et_al__2006__AA__MR_collapse_TOV,Cerda-Duran_2008,Sawai_et_al__2013__apjl__GlobalSimulationsofMagnetorotationalInstabilityintheCollapsedCoreofaMassiveStar,Sawai_Yamada_2015}.
Furthermore, it has been investigated in local shearing box
simulations
\cite{Obergaulinger_et_al_2009,Masada_et_al__2012__apj__LocalSimulationsoftheMagnetorotationalInstabilityinCore-collapseSupernovae}.
Both types of models suggest that the MRI can indeed grow under
conditions typical for PNS.  So far, these studies leave, however,
several questions open, most importantly what is the amplitude of the
turbulent stress generated by the MRI.  In a new set of simulations,
we focus on the transition from the initial phase of exponential
growth of the MRI to turbulence.  In the former phase, the MRI takes
the form of laminar \emph{channel modes}, pairs of radial inflows and
outflows containing magnetic fields of opposite polarity.  They
contain shear layers as well as current sheets, which can become
unstable against Kelvin-Helmholtz (KH) instabilities or tearing-modes
(TM).  The growth rate of these secondary, \emph{parasitic}
instabilities is proportional to the radial component of the velocity
or magnetic field, respectively.  Since these components grow
exponentially at a constant rate, $\gamma_{\mri}$, the growth rate of
the parasitic instabilities increases exponentially with time and,
consequently, their amplitude grows super-exponentially.  When the
amplitudes of the parasites are as large as the amplitude of the MRI
itself, they terminate the MRI growth, initiating the saturation
phase.  This parasitic termination scenario was described by
\cite{Goodman_Xu} and analysed in detail by \cite{Pessah}, who
determined the growth rate of the KH instabilities and TM, and
identified which of the two parasites should dominate given the
viscosity and resistivity of the gas.

To test the predictions of \cite{Pessah} under conditions
representative for the outer layers of a differentially rotating PNS,
we performed a set of local three-dimensional simulations of the MRI
in the shearing-disc approach
(\cite{Klahr_Bodenheimer__2003__ApJ__Global-baroclinic-inst-disc,Obergaulinger_et_al_2009})
varying the viscosity and resistivity.  For a detailed discussion of
our simulations, see \cite{ROCMA}.  Here, we will focus on the impact
that different numerical methods have on the results.

\section{Setup and numerical methods}
\label{Sek:PhysNum}

\begin{table}
  \centering
  \begin{tabular}{|l|ll|l|cc| }
    \hline
    name & reco. & reso. & box & $\gamma_{\mri} \, [\isek]$ & $\MMMte$
    \, $[10^{30}\ \mathrm{G}^2]$
    \\
    \hline
    PLM-8 & PLM & 8 & s & 926 & 2.2
    \\
    PLM-10 & PLM & 10 & s & 959 & 2.3
    \\
    PLM-16 & PLM & 16 & s & 1089 & 1.9
    \\
    PLM-20 & PLM & 20 & l & 1116 & 1.9
    \\
    PLM-34 & PLM & 34 & s & 1123 & 1.8
    \\
    \hline
    MP5-8 & MP5 & 8 & s & 1093 & 1.1
    \\
    MP5-10 & MP5 & 10 & s & 1104 & 1.4
    \\
    MP5-16 & MP5 & 16 & s & 1127 & 1.1
    \\
    MP5-20 & MP5 & 20 & s & 1134 & 0.85
    \\
    MP5-34 & MP5 & 34 & s & 1127 & 1.0
    \\
    \hline
    MP9-8 & MP9 & 8 & s & 1104 & 0.79
    \\
    MP9-10 & MP9 & 10 & s & 1122 & 1.3
    \\
    MP9-16 & MP9 & 16 & s & 1130 & 1.1
    \\
    MP9-20 & MP9 & 20 & l & 1126 & 1.1
    \\
    MP9-25 & MP9 & 25 & l & 1127 & 1.0
    \\
    MP9-34 & MP9 & 34 & s & 1127 & 0.93
    \\
    MP9-67 & MP9 & 67 & l & 1127 & 0.73
    \\
    MP9-134 & MP9 & 134 & s & 1128 & 0.73
    \\
    \hline
  \end{tabular}
  \caption{
    List of models.  The first column displays the name of the
    model, which is a combination of the reconstruction method
    (\second\, column) and the grid resolution, measured in grid cells
    per MRI wavelength (\third\, column).  In the \nth{4} column,
    letters indicate the size of the simulation domain: 's' (small)
    and 'l' (large)
    stand for box sizes of $1 \, \km \times 1 \, \km \times 0.333 \,
    \km$ and $1 \, \km \times 4 \, \km \times 1 \, \km$,
    respectively.  The last two columns give the growth rate of the MRI
    and the termination value of the Maxwell stress 
    (defined in Eq.\,\ref{eq:Maxwell}), respectively.
  }
  \label{Tab:models}
\end{table}

We solve the equations of visco-resistive MHD in the presence of an
external gravitational field and use the same initial and boundary
conditions as in \cite{ROCMA}.  The simulation domain covers 1\,km to
4\,km in each direction of a cylindrical box ($r, z, \phi$) centered
at a radius of $r_0 = 15.5 \, \km$.  The initial profiles of
gravitational potential, density, and pressure are derived from the
post-bounce state of the global simulations of
\cite{Obergaulinger_et_al__2006__AA__MR_collapse_TOV}.  In addition,
we assume an angular velocity with a simple power law profile,
\begin{equation}
  \Omega = \Omega_0 \left( \frac{r}{r_0} \right)^{-q},
\label{eq:omega}
\end{equation}
where $q = 1.25$, and $\Omega_0 = 1824\, \isek$.  This profile is
hydrodynamically stable.  

If we add a uniform magnetic field in $z$-direction, the MRI will grow
from small perturbations.  
The geometry and strength of the CCSN cores our models are supposed to
represent depends on the pre-collapse fields as well as the evolution
during and immediately after collapse and, as a result, are quite
uncertain.  Stellar-evolution models support a combination of poloidal
and toroidal components \cite{Braithwaite_Nordlund_2006}.  Both kinds
of fields are susceptible to the MRI.  We focus here on the case of a
poloidal, in our case vertical, field because the MRI grows fastest
for this field component.  By focusing on a uniform field, we
furthermore neglect possible small-scale structures, which might yield
small corrections to the geometry, but not the growth rate of the MRI.

We impose periodic boundary conditions in $z$ and $\phi$ direction,
and shearing-disc boundaries in radial direction, i.e.~boundaries that
are periodic in the deviation of the hydrodynamic variables from the
initial (background) state.

The parameters of our simulations are based on the models of
\cite{ROCMA}.  The theoretical growth rate of the MRI is
$\gamma_{\mri} = q \Omega_0/2 \approx 1140 \, \isek$.  We use an
initial magnetic field of $b_0^z = \zehnh{4.6}{13} \, \Gauss$,
corresponding to $\lambda_{\mri} \approx 333 \, \mathrm{m}$.  We
neglect viscosity and use a uniform resistivity
$\eta = \zehnh{4.45}{8} \, \cm \, \sek^{-1}$.  The most important
parameter that, as we will show below, causes differences between the
simulations is the grid resolution, which we express in terms of the
number of grid cells per channel width.  We present our set of models
in \tabref{Tab:models}. We point out that our simulation results are
scalable to different values of $r_0$, $\Omega_0$, and $b_0^z$ (see,
\cite{ROCMA}).

Our numerical code is based on a dimensionally unsplit finite-volume
discretisation of the MHD equations and the constrained-transport
scheme to ensure a divergence-free evolution of the magnetic field.
The spatial reconstruction can be done using the following methods:
\begin{itemize}
\item piecewise-constant reconstruction of \first order;
\item TVD piecewise-linear (PLM) reconstruction (\second order) with several
  slope limiters;
\item \nth{4}-order WENO reconstruction;
\item monotonicity-preserving reconstruction (MP;
  \cite{Suresh_Huynh__1997__JCP__MP-schemes}) of \third, \nth{5},
  \nth{7}, or \nth{9} order.
\end{itemize}
In \cite{ROCMA}, we used the most accurate of these schemes,
viz.~\nth{9}-order MP reconstruction.  Here, we will make a comparison
to two methods of lower order: \nth{5}-order MP, and PLM with the MC
slope limiter.  The other ingredients of the code are the same as in
\cite{ROCMA}: a multi-stage (MUSTA) version of the HLLD Riemann solver
and a \third-order Runge-Kutta time integrator.

\section{Results}
\label{Sek:Res}

For the conditions in supernova cores, \cite{ROCMA} found, as
predicted by \cite{Pessah}, that the MRI growth is terminated by KH
modes, unless special conditions such as an insufficient size of the
numerical domain somewhat artificially restrict their growth.  This
qualitative finding also holds for less accurate reconstruction
schemes.

\begin{figure}
  \centering
  \includegraphics[width=\linewidth]{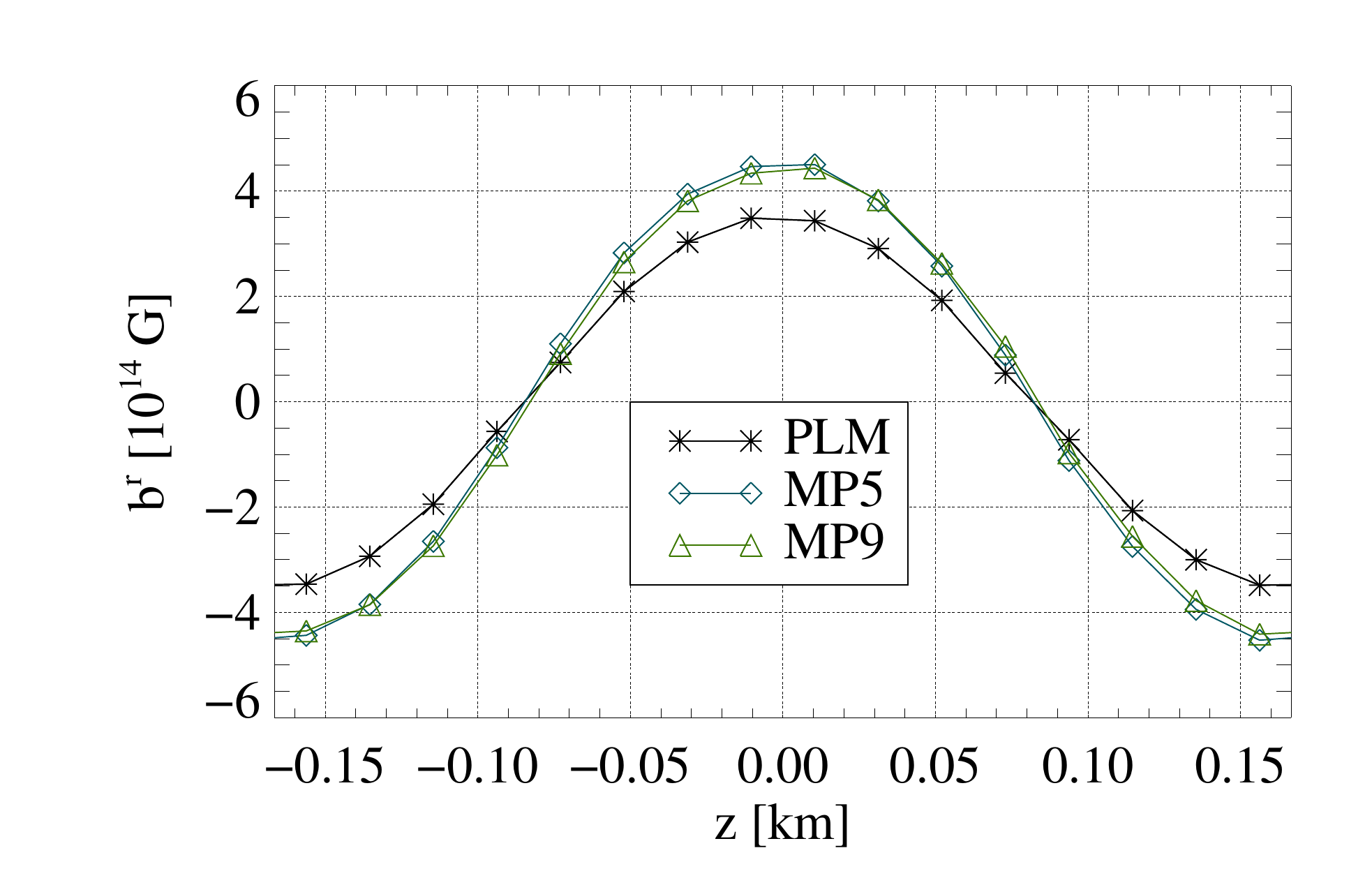}
  \caption{
    Vertical profiles of the radially averaged radial magnetic field 
    at $t = 10 \, \msec$ for models PLM-16, MP5-16, and MP9-16.
  }
  \label{Fig:channelprofile}
\end{figure}

Irrespective of the reconstruction method, MRI channel modes start to
grow from the weak perturbations introduced at $t = 0$.  Even for low
grid resolution and low-order schemes, the vertical profiles that we
find in the simulations approximate very well sine functions, as we
show for the radial component of the magnetic field of models PLM-16,
MP5-16, and MP9-16 in \figref{Fig:channelprofile}.  The numerical
growth rates of the MRI depend on the resolution and the
reconstruction scheme.  Under-resolved simulations show in general a
slower MRI growth, but as the grid resolution increases, all methods
exhibit numerical convergence to the theoretical value.
The good agreement between numerical schemes shown in
\figref{Fig:channelprofile} holds well throughout the exponential
growth phase.  Differences show up when the growth is terminated,
leading to a spread in the value of the termination amplitude (see
below).

\begin{figure}
  \centering
  \includegraphics[width=0.48\textwidth]{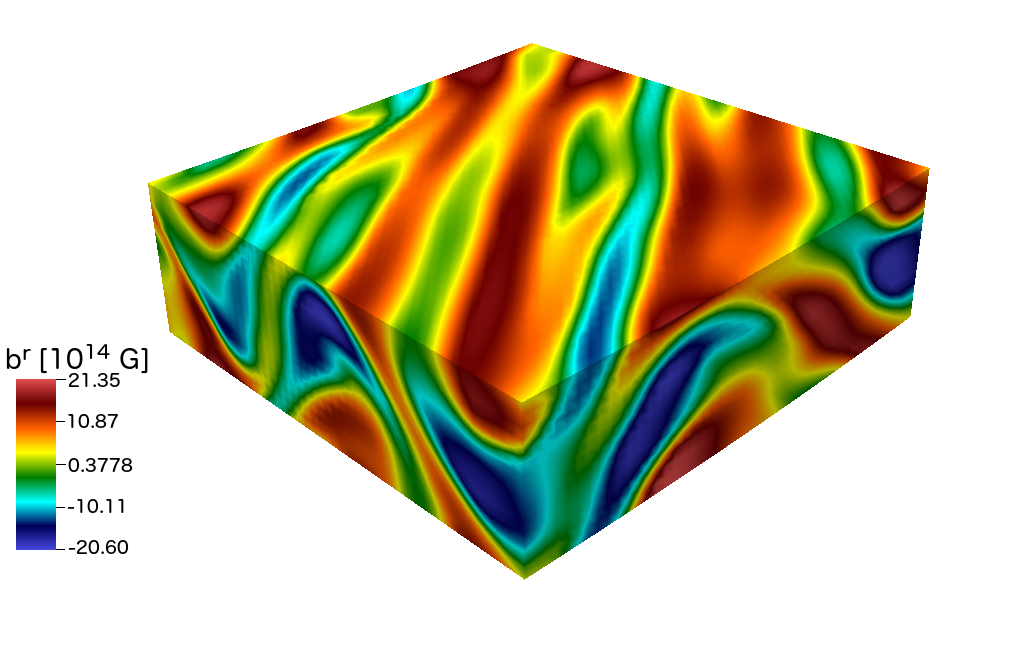}
  \includegraphics[width=0.48\textwidth]{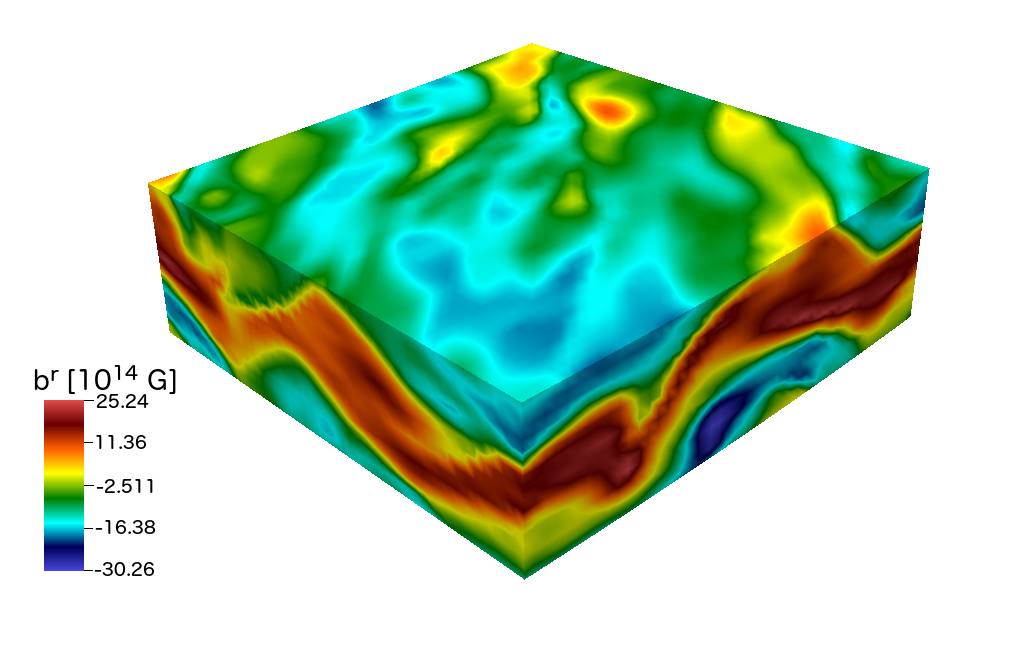}
  \includegraphics[width=0.48\textwidth]{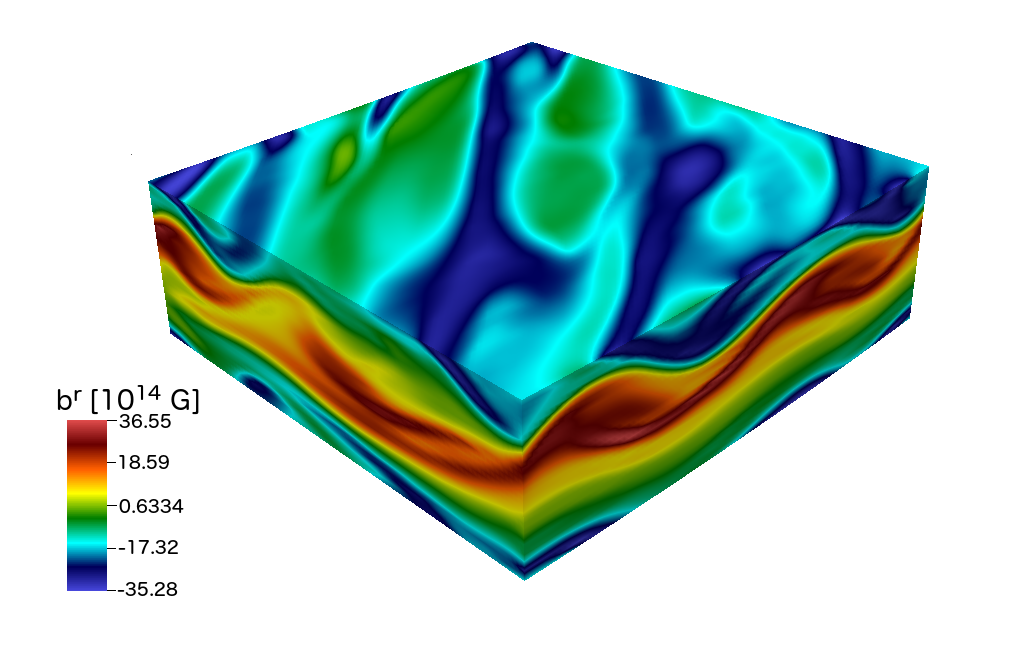}
  \includegraphics[width=0.48\textwidth]{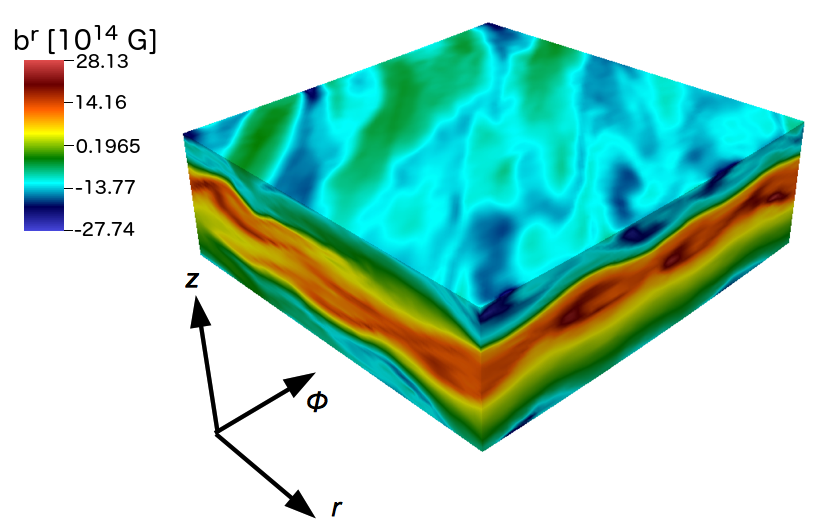}
  \caption{
    Visualisation of the radial component of the magnetic field close
    to the moment of termination for models (top left to bottom right)
    PLM-16 (at $t = 12\, \msec$), MP9-16 (at $t = 11.5\,\msec$),
    PLM-34 (at $t = 11.5\, \msec$), and MP9-34 (at $t = 11\,
    \msec$), respectively. Note that not all snapshots are taken at the same
    evolutionary time, but they correspond to the same evolutionary
    instant, namely the time at which (for each model) saturation of
    the MRI growth sets in.
  }
  \label{Fig:3dplots}
\end{figure}

The channel modes constitute the background in which parasitic
instabilities grow.  We compare visualisations of the radial component
of the magnetic field for models with PLM and MP9 reconstruction
schemes and two resolutions in \figref{Fig:3dplots}.  We note that the
snapshots show a very dynamical phase of the evolution, during which a
slight time difference corresponds to a very pronounced difference in
the state of the flow.  Therefore (and because of the influence of
random perturbations), a point-wise comparison between the four panels
cannot be made.  Nevertheless, we conclude that the parasites have the
same structure as the one found in \cite{ROCMA}, and thus can be
identified as KH instabilities.  In particular, patterns such as the
vortices visible in the $r\text{--}z$-planes of the lower two panels
provide evidence for KH modes.  The rather different appearance of the
two upper panels (with a resolution of 16 zones per channel) is not
caused by a difference in the parasites, but is a consequence of the
fact that the parasites have grown somewhat more than in the two
bottom panels.

Analogously to MRI channels, KH modes grow at a reduced rate if the
numerical resolution is insufficient.  This means that, for the same
MRI amplitude, the KH parasites grow slower. Consequently, the time at
which the parasites are strong enough to terminate the MRI growth is
delayed for a coarse grid.  Furthermore, KH parasite modes develop
structures smaller than the channel mode.  Therefore, insufficient
resolution affects them more than the MRI, and parasitic modes can be
suppressed on a grid where the MRI growth rate is hardly reduced.
Thus, non-converged simulations tend to \emph{overestimate} the
termination amplitude of the MRI as measured by, e.g.~the maximum
value of the volume-averaged Maxwell stress component,
\begin{equation}
  \label{eq:Maxwell}
  \MMM =   \frac{  | \int b^r b^{\phi} \mathrm{d} V | } {V}, 
\end{equation}
where $V$ is the total grid volume.  This tendency is ameliorated for
more accurate numerical schemes.

\begin{figure}
  \centering
  \includegraphics[width=0.32\textwidth]{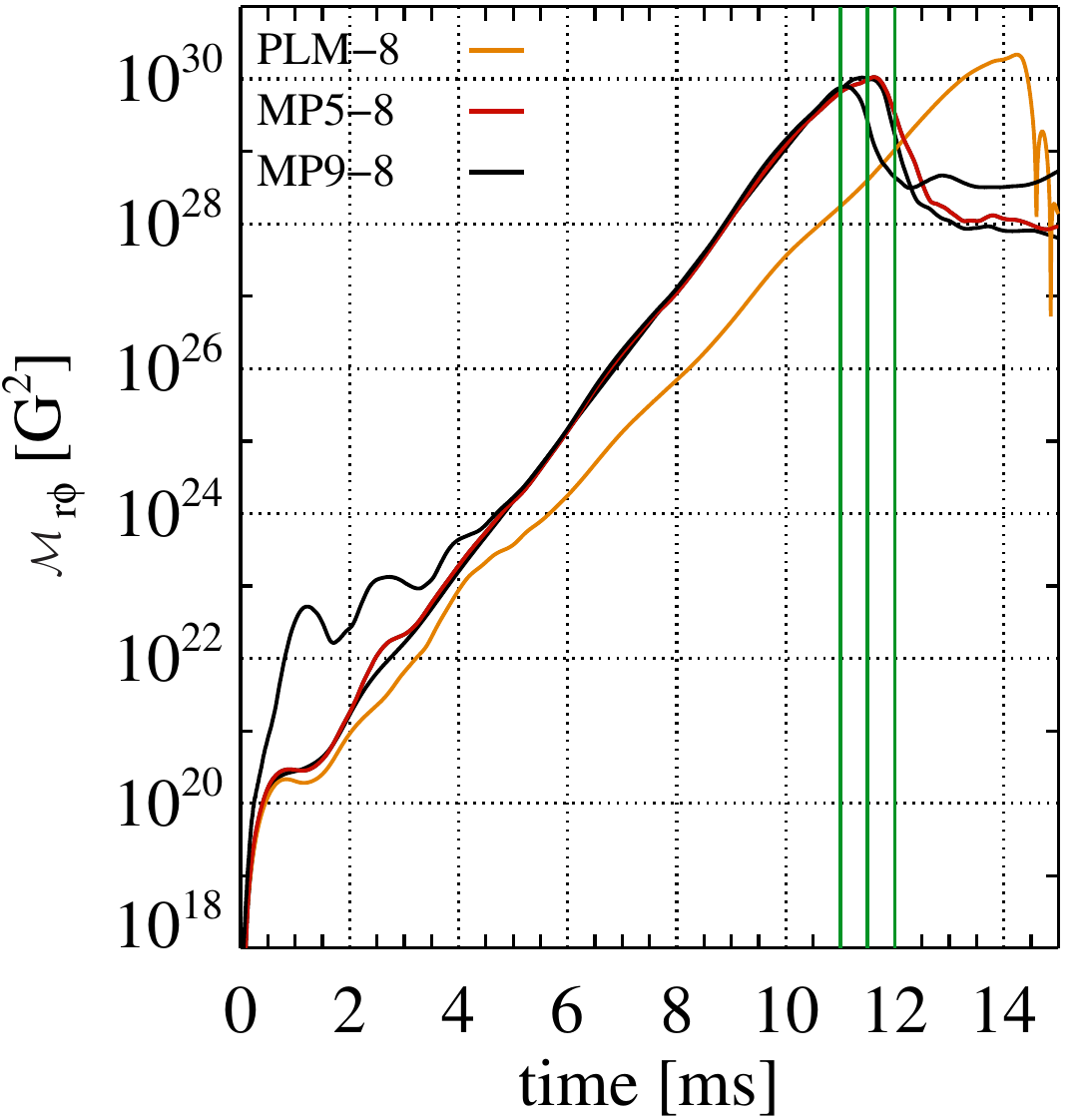}
  \includegraphics[width=0.32\textwidth]{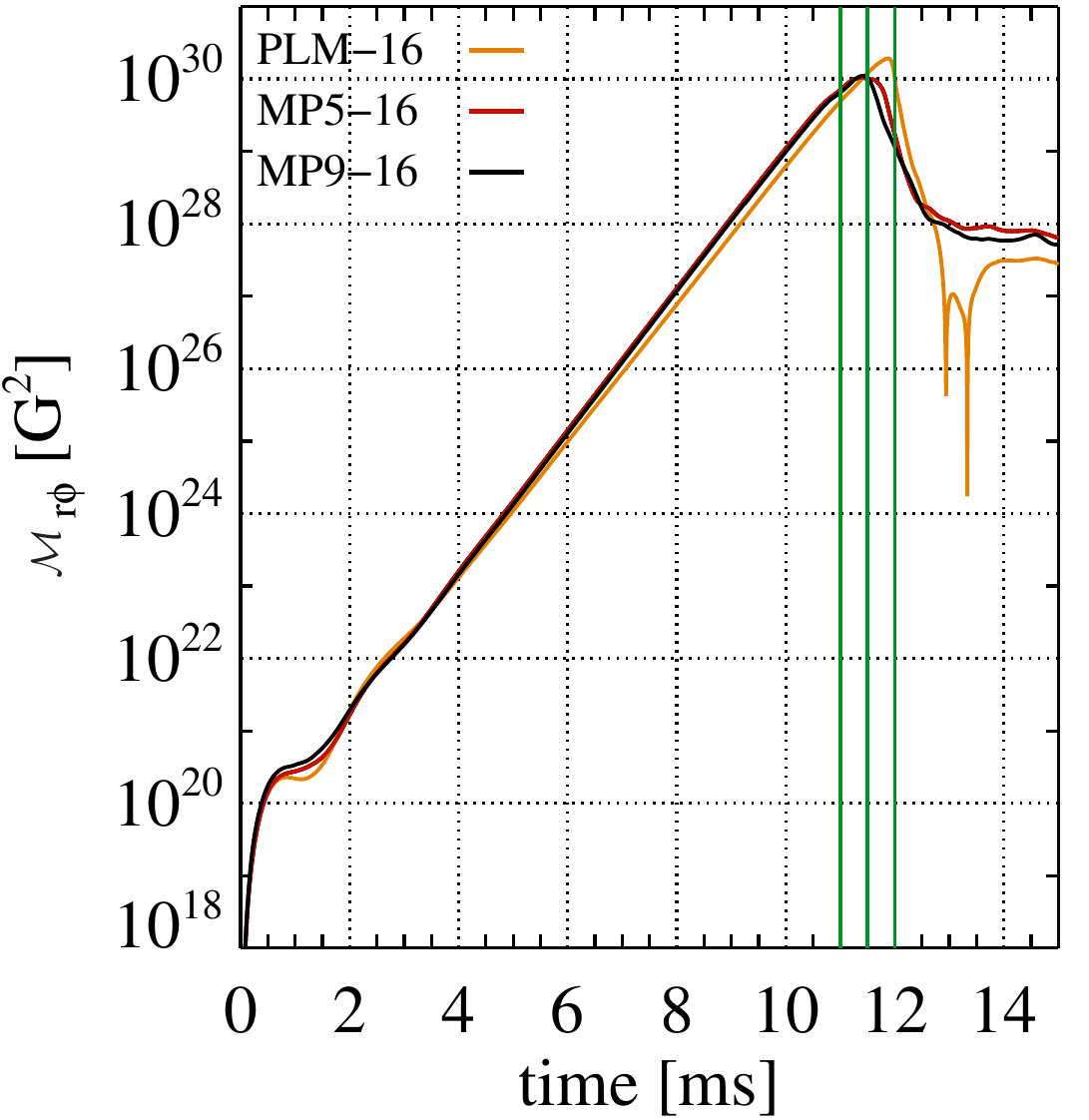}
  \includegraphics[width=0.32\textwidth]{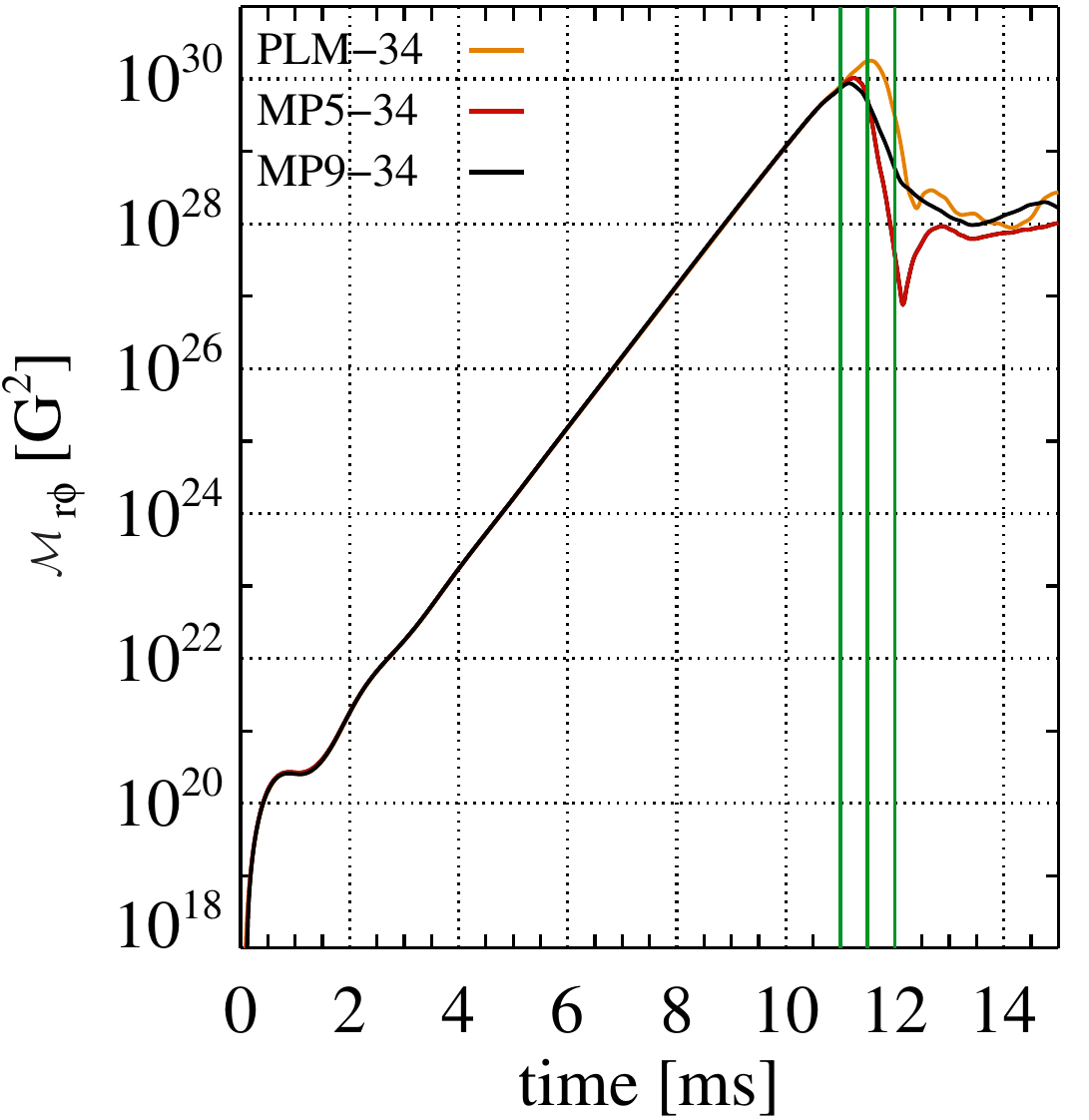}
  \includegraphics[width=0.48\textwidth]{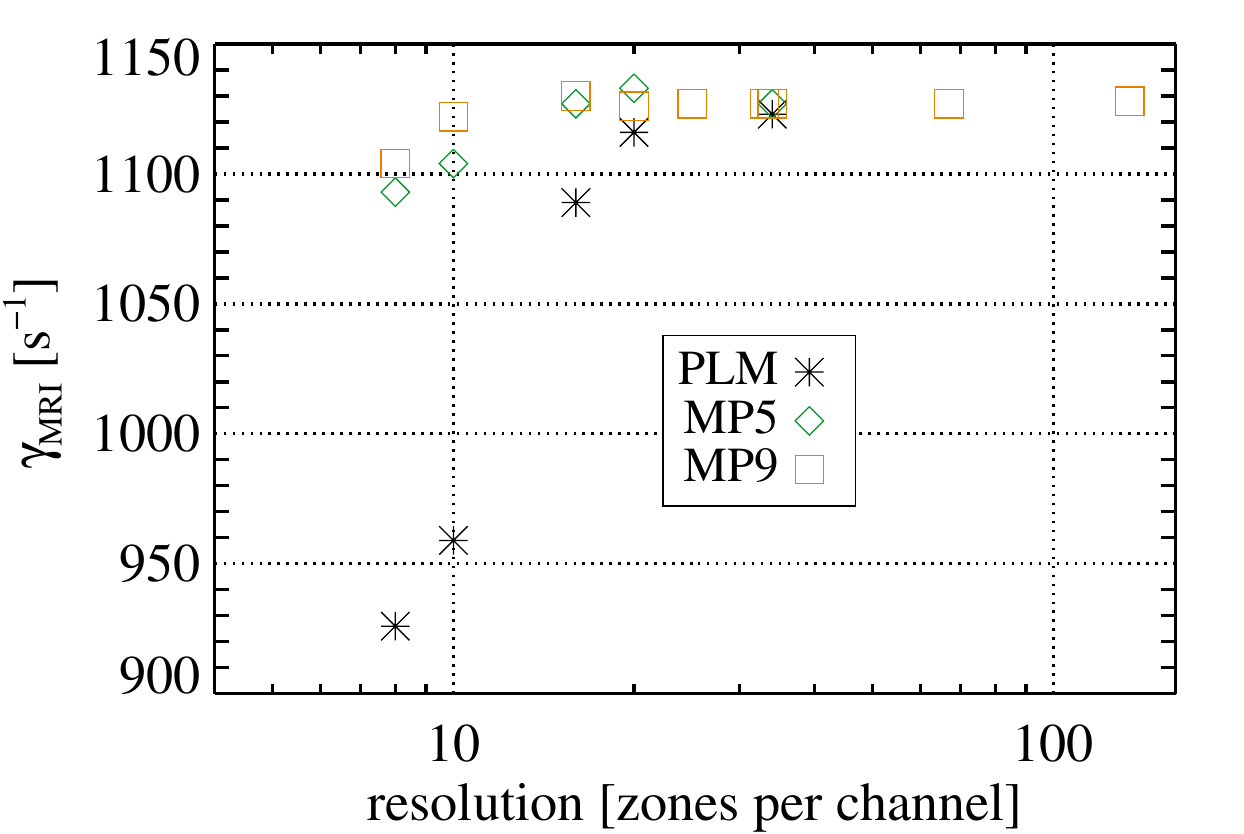}
  \includegraphics[width=0.48\textwidth]{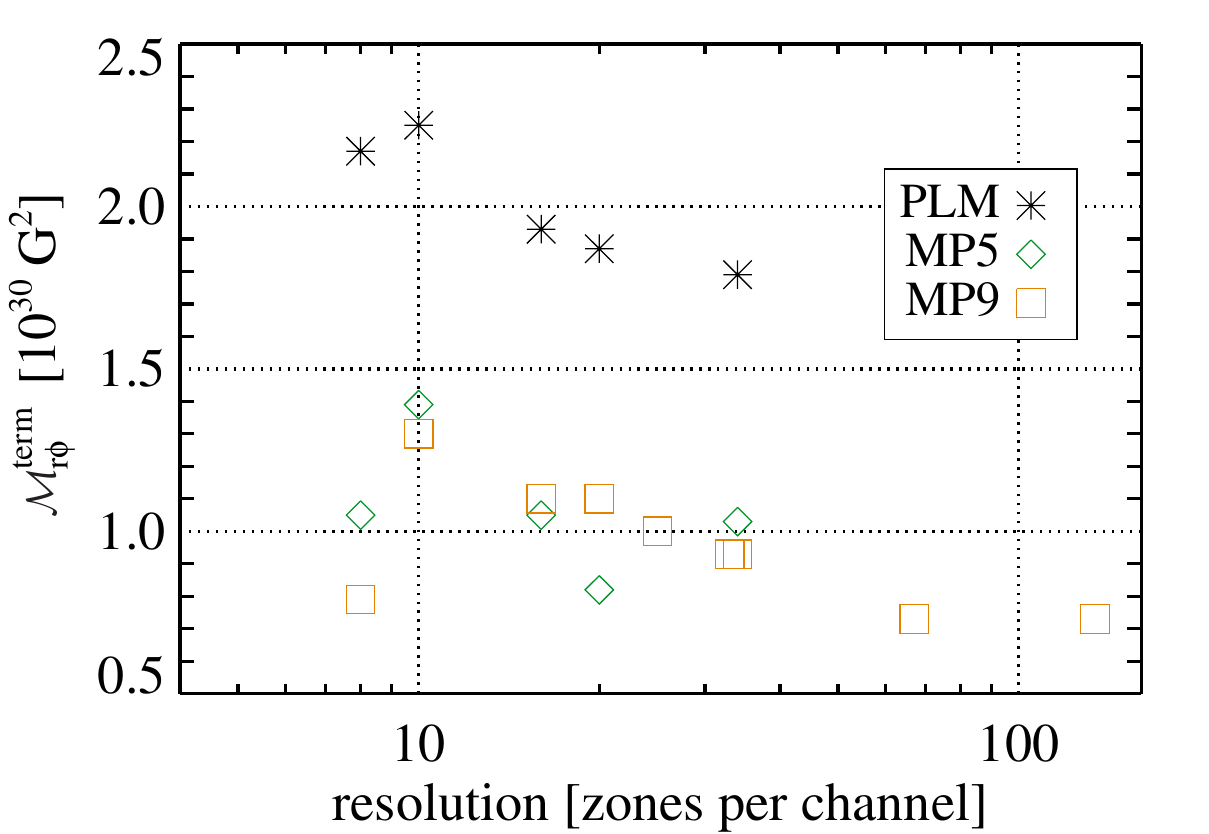}
  \caption{
    Top panels: time evolution of $\MMM$ for models with grid resolutions
    of 8, 16, and 34 zones per channel (left to right).  Bottom
    panels: dependence of  the growth rate (left)  and the
    termination amplitude (right) of the MRI on the grid resolution.
  }
  \label{Fig:reso}
\end{figure}

We summarise the results described above in \figref{Fig:reso}.  The
three top panels show the time evolution of $\MMM$ for simulations
with resolutions of 8, 16, and 34 zones per channel.  Except at the
lowest resolution, for which the PLM run exhibits a low growth rate,
the exponential MRI growth is captured well.  Comparing the growth
rates for all simulations (bottom left panel), we can quantify the
differences between the reconstruction methods: PLM requires roughly
34 zones per channel to attain a growth rate $\gamma_{\mri} > 1120 \,
\isek$, whereas the MP schemes achieve that value already at 16 (MP5)
and 10 (MP9) zones, respectively.

In the top panels, MRI termination corresponds to the maximum value of
$\MMM(t)$.  The bottom right panel compares this maximum for all
models.  Because of the suppression of the growth of the parasites at
low resolutions, the termination amplitude increases with resolution
until a sufficiently large number of zones per channel width is
reached. Then it settles to a roughly constant value.  Please note
that the runs with 8 zones per channel represent a special case for
which other effects beside parasitic KH modes affect termination such
as the diffusion of magnetic field and momentum across the current and
shear layers, respectively.  Furthermore, model MP5-20 deviates from
the trend, possibly indicating the range of stochastic variations of
the termination amplitude.  We find considerable differences between
PLM and the two MP schemes, with the former yielding an almost twice
as high $\MMMte$ as the two other at the same resolution.
Restrictions in computing time allowed us to obtain numerical
convergence only for the MP9 models, which approach a value of $\MMMte
= \zehnh{0.73}{30} \, \mathrm{G}^2$, if we use at least 67 zones per
channel.  We expect, however, the other models to behave similarly for
a sufficiently high grid resolution.

\section{Summary and conclusions}
\label{Sek:Sum}

In an extension of the studies of \cite{Obergaulinger_et_al_2009} and
\cite{ROCMA}, we investigated the influence of numerical effects on
the early evolution of the MRI in conditions similar to those expected
in core-collapse supernovae.  We performed a series of
three-dimensional MHD simulations in the shearing-disc framework,
fixing the physical conditions (rotational profile, density, initial
field) in such a way that we could observe the growth of MRI channel
modes and their disruption by parasitic Kelvin-Helmholtz modes and
varying the grid resolution and the reconstruction method.  For the
latter, we used a \second\, order piecewise linear reconstruction with
the MC slope limiter, as well as monotonicity-preserving methods of
\nth{5} and \nth{9} order accuracy.

The main results of this study are the following:
\begin{itemize}
\item The physical nature of the parasitic instability terminating the
  MRI does not depend on the numerical settings.  The structure of the
  flow and the field show the signatures of KH modes, which are
  consistent with the theoretical predictions of \cite{Pessah}.
\item All methods capture the exponential growth of the MRI well at
  low resolution of the order of 10 zones per channel, with MP schemes
  performing better than PLM.
\item The termination value of the Maxwell stress exhibits convergence
  with grid resolution.  Under-resolved simulations overestimate its
  value because numerical viscosity reduces the growth rate of the
  parasites.  Again, MP schemes are superior to PLM.  In general, a
  much higher resolution is required to obtain the correct parasitic
  growth rate than that of the MRI itself, because KH modes grow on
  much smaller length scale, viz.~the extent of the shear layer
  between the channel flows.
\end{itemize}


From these results we conclude that the most important findings
concerning the MRI termination are robust w.r.t.~the effects of
numerical schemes, which affirms our previous studies.  Our initial
field strengths, and thus the MRI wave lengths, most likely exceed
those that can be expected in most supernova cores (the
  post-collapse fields of \cite{OJA_2014} would, in a rapidly rotating
  core, correspond to $\lambda_{\mri} \sim 0.01...10 \, \mathrm{m}$).
This means that for realistic initial fields, the constraint to
resolve the KH instability in the shear layer can translate into a
very severe restriction on the grid resolution beyond what can be
currently afforded in global models.
Simulations that cannot meet this resolution requirement run the risk
of producing too high amplitudes of the MRI at termination, which may
lead to unphysically strong magnetic stresses and consequently tend to
overestimate the influence of magnetic fields on the dynamics.
Although the use of high-order methods, such as our MP5 and MP9
schemes, does not overcome this limitation, at least it helps to
alleviate it.  We estimate that a three-dimensional simulation with a
high-order method following the evolution for a time scale of $\sim
100\,$ms and resolving channels of $1\,$m length scale in a PNS of
radius $30\,$km requires $\sim 10^{8}$ grid cells and $\sim 10^{14}$
CPU hours.  Even if adaptive mesh refinement (AMR) techniques were
employed, the simulations would still be beyond current possibilities.

\section*{Acknowledgments}

TR acknowledges support from The International Max Planck Research
School on Astrophysics at the Ludwig Maximilian University Munich, EM
\& TM acknowledge support from the Max-Planck-Princeton Center for
Plasma Physics, and MA, PCD, TR and MO acknowledge support from the
European Research Council (grant CAMAP-259276). We also acknowledge
support from grants AYA2013-40979-P and PROMETEOII/2014-069.  The
computations have been performed at the Leibniz Supercomputing Center
of the Bavarian Academy of Sciences and Humanities (LRZ), the Max
Planck Computing \& Data Facility (MPCDF), and at the Servei
d'Inform\`atica of the University of Valencia.

\bibliographystyle{mn2e}

\end{document}